\documentclass[11pt]{article}

\usepackage[dvips]{color, graphicx}

\newcounter{pubnum}
   {\begin{list}{\arabic{pubnum}}%
       {\usecounter{pubnum} \itemsep 0.1in}}
   {\end{list}}

\title{Gravitational Waves: new observatories for new astronomy}

\author{Louis J. Rubbo, Shane L. Larson, Michelle B. Larson, and 
        Kristina D. Zaleski\\
	Center for Gravitational Wave Physics, The Pennsylvania
        State University\\
	University Park, PA 16802}

\begin{document}


\maketitle


\section{Introduction}

Nearly ninety years ago, Albert Einstein made a startling discovery
with general relativity, his newly published relativistic description
of gravity: gravity can propagate in waves and carry information from
one place in the Universe to another, just like photons.  It was
simple to calculate that these waves would be very weak, and they were
generally dismissed as being unobservable and thus unimportant as an
experimental probe of general relativity.  Other experiments, such as
the deflection of starlight, were easily detectable with early 20th
century technology, and helped pave the way for the acceptance of
general relativity as the correct, modern description of gravity.

Now, almost a century later, technology has caught up with theory, and
the detection of gravitational waves is no longer an implausible
dream.  Large gravitational-wave observatories are being built to
listen for faint echoes of gravitational waves that have propagated
across the far reaches of the Universe to us here on Earth.  The
detection of gravitational waves from dynamic astrophysical systems
will provide astronomers with unprecedented information about the
Universe, yielding information that is unaccessible using traditional
electromagnetic observations.  Gravitational waves will allow the
detailed study of dark compact objects, such as black holes, or the
interiors of supernova, or possibly the formation of the Universe
itself.

This article reviews the current status of gravitational wave
astronomy and explains why astronomers are excited about the new
generation of gravitational wave detectors.  As part of the review we
compare and contrast gravitational radiation to the more familiar
electromagnetic radiation.  We discuss the current indirect experimental
evidence for gravitational waves, and how current and future
gravitational wave detectors will operate as our newest telescopes
are pointed at the skies.


\section{Multi-spectrum Astronomy}

Electromagnetic astronomy makes leaps and bounds in sensitivity and
angular resolution by constructing new, more capable observatories on
the ground and in space.  New technology, such as adaptive optics and
interferometry, help advance our ability to observe the Cosmos.
However, despite our best efforts, there are regions of the Universe
that will forever remain shrouded to electromagnetic telescopes
because matter impedes the propagation of light to our detectors on or
near the Earth.

In some instances, simply changing the wavelength of light allows us
to see through intervening matter.  This is the case with the Milky
Way where vast clouds of dust and gas in the plane of the galaxy block
visible light emission from the galactic center, but infrared light is
able to get through.  There are other cases, however, where photons of
any wavelength cannot escape because the matter is too dense.  At the
core of a supernova, beneath the collapsing envelope of the stellar
atmosphere, photons are trapped by the sheer density of matter.  Some
time after the star explodes, and the expanding shell of gas is thrown
off, the remnants of the star expand to a less dense state.  At this
point the photons find themselves free to travel through space and be
received on Earth as messengers from the explosion.  A similar fate
befalls photons originating within the first 300,000 years after the
Big Bang.  During this era the Universe was also very dense, and no
form of light could propagate freely.  Every photon was perpetually
tangled in a sea of dense matter, bouncing around like a fly in a maze
of window panes.  As the Universe expanded it became less dense, and
eventually the photons were able to propagate freely.  The transition
point is known as the ``recombination curtain'', and the free
streaming photons associated with this era are collectively known as
the Cosmic Microwave Background.

These kinds of dense environments define the limits of our vision
using electromagnetic telescopes.  Photons will never arrive from the
core of a supernova or from a time before the recombination curtain.
It would be great if there was a way to see into these super dense
regions, analogous to seeing the center of the galaxy in infrared.
Fortunately there is: we can look in gravitational waves.

Unlike electromagnetic radiation, gravitational radiation interacts
weakly with matter.  Gravitational waves propagate through space
freely, taking little notice of the environments they pass through.
Gravitational waves are generated by dynamic and energetic motion of
matter, as one might find in the core of a collapsing star or in the
very early Universe.  Most importantly, they carry information about
times and places in the Cosmos that we have no other ways to probe.
Detecting gravitational radiation will reveal for the first time the
secrets of environments which have to date been the realm of
theoretical calculations and speculation.

In addition to probing environments too dense for light to escape,
gravitational waves will provide insight into astronomical systems too
compact and dim to resolve with current telescopes.  Observations of
bumps on neutron stars, stellar mass objects orbiting super-massive
black holes, or the final merger of two stellar remnants are all well
outside the capabilities of even the most advanced electromagnetic
telescopes.  By contrast, these systems should be detectable with high
confidence in gravitational waves, and will provide information that
is unobtainable by any other means.  

\section{New Eyes on the Universe}

Photons love interacting with matter.  On the up-side, this makes them
easy to gather with telescopes (made of matter), but on the down-side,
photons are easily distracted by intervening matter during their
flight to Earth.  In contrast, gravitational waves interact very
poorly with matter.  As a result, gravitational waves make a great
probe of the Universe, but they present a significant challenge to
designing an astronomical instrument (made of matter) which can sense
their passing.

We have strong indirect evidence for the existence of gravitational
waves from electromagnetic observations of the Hulse-Taylor binary
pulsar and others like it.  In the early 1970's, the team of Russell
Hulse and Joseph Taylor used radio observations to measure a decrease
in the orbital period for the binary pulsar PSR 1913+16.  Subsequently
they showed that the energy loss associated with the orbital decay
rate was consistent with the emission of gravitational waves.  For
their discovery, Hulse and Taylor won the 1993 Nobel Prize in physics.

Indirect evidence is not the same as having direct observations, and a
direct detection of gravitational waves has not yet been made.  If we
are to detect gravitational waves directly, we need to first
understand what must be measured.  Gravitational waves are
oscillations that stretch and squeeze the fabric of spacetime in a
characteristic way that reflects the motion of the emitting system.
Similar to electromagnetic radiation, gravitational waves come in two
flavors, or {\em polarizations}.  These polarization states are
referred to as \textit{plus} ($+$) and \textit{cross} ($\times$) after
the pattern of stretching and squeezing they impose on matter that
they pass through.  The gravitational waves bathing the Earth are
expected to be extremely weak.  A strong gravitational wave is only
expected to stretch (or squeeze) spacetime by the width of an atomic
nucleus over a distance of five million kilometers!  Detecting its
presence will be like looking for an atom-size change in distance
between the ends of a ruler thirteen times as long as the distance
between the Earth and the Moon!

Direct detection of something as weak as a gravitational wave requires
a not-so-conventional observatory.  Gravitational wave detectors fall
into two broad categories: resonant bar and interferometric detectors.
Bar detectors listen for gravitational waves by monitoring acoustic
modes in an isolated metallic bar.  When a gravitational wave with a
frequency similar to the resonant frequency of the bar stretches and
squeezes the bar, it will ``ring''.  Joseph Weber (1919 - 2000)
brought the first bar detector online in the 1960's and today there
are several operating around the world.  In Table~\ref{tab:detectors}
we present a representative list of detectors currently in use.
\begin{table}[tb]
  \caption{A representative list of gravitational wave detectors
  either operational now or in their developmental phase.}
  \begin{center}
    \begin{tabular}{lllc}
      \hline\hline
      Detector & Type & Location & Operational \\
      \hline
      EXPLORER & Resonant Bar & Geneva, Switzerland & 1984 \\
      ALLEGRO & Resonant Bar & Baton Rouge, LA, USA & 1986 \\
      NAUTILUS & Resonant Bar & Rome, Italy & 1994 \\
      AURIGA & Resonant Bar & Lengaro, Italy & 1997 \\
      TAMA & Interferometer & Mitaka, Japan & 1999\\
      LIGO & Interferometer & Livingston, LA, USA & 2002 \\
      & & Hanford, WA, USA & \\
      GEO 600 & Interferometer & Hannover, Germany & 2002 \\
      VIRGO & Interferometer & Cascina, Italy & 2005 \\
      LISA & Interferometer & Space & $\sim$2014 \\
      \hline\hline
    \end{tabular}
  \end{center}
  \label{tab:detectors}
\end{table}

Interferometric detectors operate on the principle that a
gravitational wave stretches and squeezes spacetime, changing the
proper distance between two fixed masses when it passes by. A passing
gravitational wave is detected by noting a change in the interference
pattern created when laser light from two different arms of the
interferometer is recombined. The interference pattern change occurs
when the light travel time of the laser in one arm changes relative to
the other. In the United States, the Laser Interferometer
Gravitational-wave Observatory (LIGO) operates two sites, a 4 km arm
interferometer in Livingston, LA, and two interferometers (4 km and 2
km) in Hanford, WA. Several interferometric gravitational wave
detectors are now in operation (see Table~\ref{tab:detectors}) and
currently approaching sensitivities that should make a direct
detection in the near future.

Work is also progressing on a space-based mission to detect
gravitational waves in a lower frequency range than ground based
detectors. The Laser Interferometer Space Antenna (LISA) is a joint NASA-ESA mission
currently scheduled to launch around 2014. LISA will operate as an
interferometric gravitational wave detector, but with arms that are
five million kilometers long!

Different gravitational wave detectors will detect sources in different
gravitational wave frequency bands. LIGO is sensitive to high frequency
gravitational waves (around 1000 Hertz, with wavelengths of 300 km), while
LISA will be sensitive to lower frequency gravitational radiation (around
1 milliHertz or wavelenghts of 300 million km). In this way, gravitational
wave detectors will probe different aspects of the universe, much like
traditional telescopes do when observing, for example, in infrared or radio.

\section{Gravitational Wave Astronomy}

Gravitational wave astronomy is a new and vibrant field of
observational astronomy, just entering an era where detectors capable
of detecting signals from space are becoming operational.  At the end
of 2005, the LIGO detectors will begin their first year long
science run at their target design sensitivity, giving us our first
deep probe of the Cosmos in gravitational waves.  Our expectation is
that the deep reaches of space are alive with gravitational wave
signals from black holes, neutron stars, supernovae and other highly
energetic events, and within the next decade we will be detecting them
on a regular basis.

The current generation of gravitational wave observatories is only
our first step toward looking at the Universe with new eyes and ears,
in much the same spirit as Grote Reber's first radio antenna or Isaac
Newton's first reflecting telescope.  As with all branches of
astronomy, new windows on the Universe resolve old debates about the
nature of distant astrophysical systems, and pose new mysteries and
questions about the Cosmos.  Gravitational wave astronomy is expected
to be no different.

\bigskip

\noindent {\large \textbf{Select Introductory Gravitational Wave
    Resources}}

\begin{itemize}
\item \textit{Einstein's Unfinished Symphony: Listening to the Sounds
  of Space-Time}, Marcia Bartusiak, Joseph Henry Press (2000), Berkley
  Books (2003).

\item \textit{Ripples in Spacetime}, W. Wayt Gibbs, Scientific
  American, April 2002.

\item \textit{Catch a Gravitational Wave}, Marcia Bartusiak,
  Astronomy, October 2000.

\item \textit{Teaching Einstein to Dance: The Dynamic World of General
  Relativity}, Adam Frank, Sky and Telescope, October 2000.

\item \textit{A Prehistory of Gravitational Waves}, Daniel Kennefick,
  Sky and Telescope, October 2000.

\item \textit{LIGO: An Antenna Tuned to the Songs of Gravity}, Gary
  H. Sanders and David Beckett, Sky and Telescope, October 2000.
\end{itemize}


\bigskip

\noindent This work was supported by the Center for Gravitational Wave
Physics (NSF) grant PHY-01-14375.



\end{document}